\documentclass[a4paper]{article}
\pdfoutput=1
\usepackage{icrc2013}
\usepackage[english]{babel}

\title{In-situ measurements of whole-dish reflectivity for VERITAS}

\shorttitle{Whole-dish Reflectivity}

\authors{
Simon Archambault$^{1}$,
Sean Griffin$^{1}$,
and David Hanna$^{1}$
}

\afiliations{
$^1$ Department of Physics \\
McGill University \\
Montreal, QC H3A 2T8, Canada}

\email{archs@physics.mcgill.ca}

\abstract{The VERITAS array is a set of four imaging atmospheric Cherenkov telescopes (IACTs) sensitive to gamma rays at energies between 85 GeV and 30 TeV. Each telescope is based on a tessellated mirror, 12 metres in diameter, which reflects light from a gamma-ray-induced air shower to form an image on a pixellated `camera' comprising 499 photomultiplier tubes. The image brightness is the primary measure of the gamma ray's energy so a knowledge of the mirror reflectivity is important. We describe here a method, pioneered by members of the MAGIC collaboration, to measure the whole-dish reflectivity, quickly and regularly, so that effects of aging can be monitored. A CCD camera attached near the centre of the dish simultaneously acquires an image of both a target star and its reflection on a target of Spectralon, a highly-reflective material, placed at the focus of the telescope. The ratio of their brightnesses, as recorded by the CCD, along with geometric factors, provides an estimate of the dish reflectivity with few systematic errors. A filter wheel is placed in front of the CCD camera, allowing to measure the reflectivity as a function of wavelength. We present initial results obtained with the VERITAS telescopes during 2012.}

\keywords{Cherenkov Telescopes, Calibration, Reflectivity}

\begin{document}
\maketitle

\section{Introduction}

Very-high-energy (VHE) gamma-ray astronomy makes use of arrays of imaging
atmospheric Cherenkov telescopes (IACTs).
A telescope consists of a large tesseleted mirror that focusses Cherenkov light 
from extensive air showers onto a camera comprising an array of photomultiplier
tubes (PMTs).
The summed signals from the PMTs is proportional to the energy of the  
incident gamma ray so a key parameter needed to 
extract science from the data acquired using such 
telescopes is the effective reflectivity of the mirror, the
``whole-dish'' reflectivity.
This number can be estimated using measurements on individual
facets made with a portable, commercially available, device and 
corrections resulting from the shadowing effects of the camera and its
support structure can be introduced. 
However it is desirable to have alternative ways of determining the 
mirror reflectivity if only to build confidence in one's understanding of the
instrument and as a way of estimating systematic errors.

In this report we describe our experience with a system developed to measure
the whole-dish reflectivity of the VERITAS telescopes. 
The method we follow was first suggested by members of the MAGIC 
collaboration~\cite{magic1,magic2}.
The basic idea is to mount a digital camera on the telescope to
record, in the same image, 
light coming directly from a bright star as well as the light
from the star that has reflected off the main mirror and subsequently off a
target of known reflectivity placed at the focus. 
Up to numerical factors, the whole-dish reflectivity is determined from 
the ratio of the two signals.
The use of a single camera to simultaneously record both the direct and 
reflected images eliminates many possible systematic errors.

\section{The VERITAS Reflectometer}

VERITAS comprises an array of four IACTs located at the Whipple Observatory 
on Mount Hopkins in southern Arizona~\cite{holder, weekes}.
Each of the telescopes is based on a 12-m diameter Davies-Cotton reflector 
focussing light onto a 499-pixel camera made from close-packed 
Hamamatsu R10560 PMTs coupled to light concentrators.  
The reflector is made up of 345 identical mirror facets; when they are 
perfectly aligned 
the on-axis point-spread-function is smaller than a pixel 
diameter.

\begin{figure}[h]
  \centering
  \includegraphics[width=0.45\textwidth]{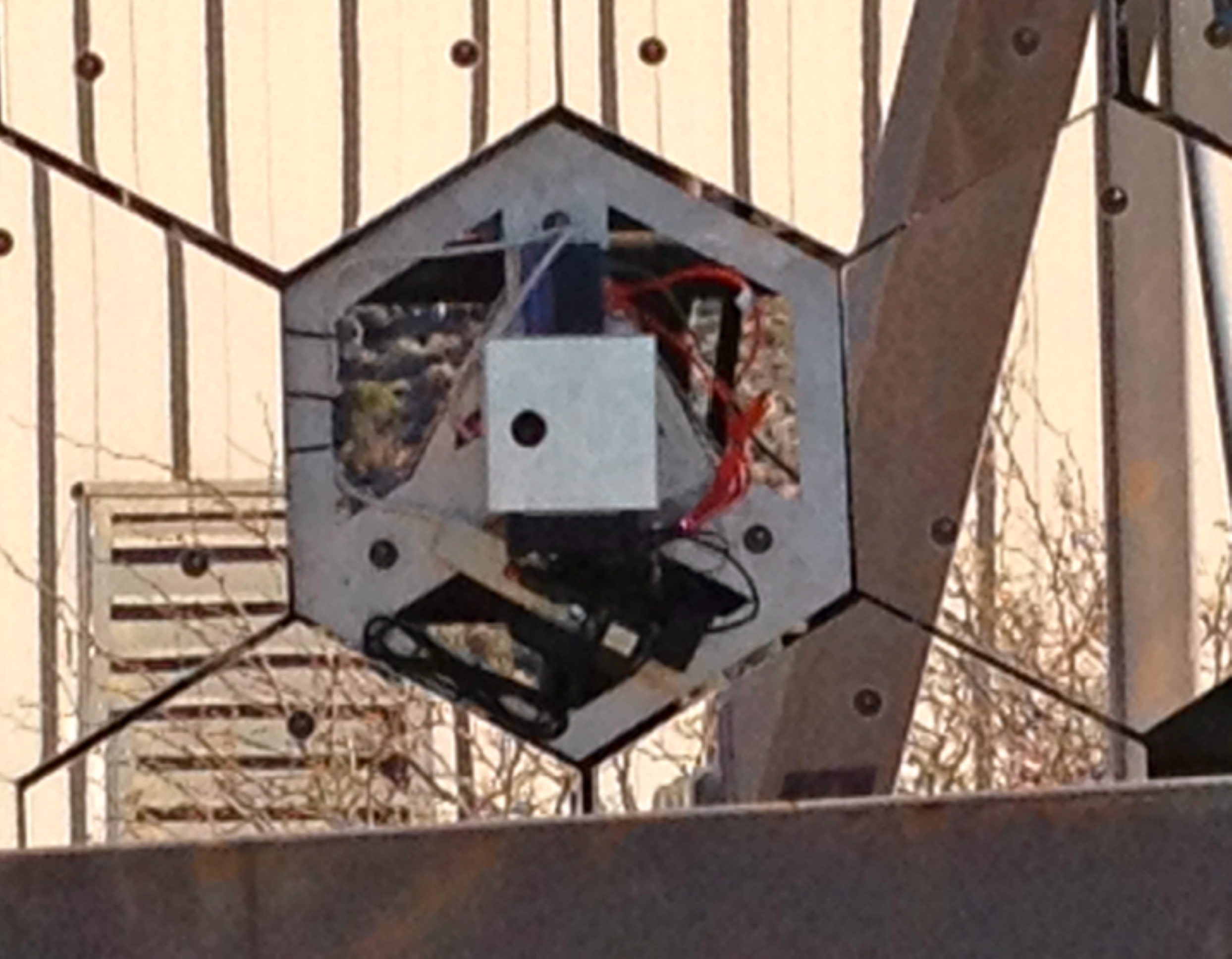}
  \vskip 0cm
  \caption{The reflectometer box is mounted on the VERITAS optical support 
structure using the same mounting scheme as the mirror facets. 
This allows easy adjustment of the pointing direction of the CCD camera.}
  \label{box-mount}
 \end{figure}

The digital camera used for this work is a Prosilica GC1380 CCD camera 
equipped with a 25 mm Pentax C-mount lens.
The camera was chosen primarily for cost and convenience reasons; we use the 
same camera as that which forms
part of a tool developed for precisely aligning the mirror 
facets~\cite{mccann}.
In order to study possible wavelength-dependent effects we employ a filter 
wheel equipped with band-pass filters. 
The filters have transmission curves 10 nm wide (FWHM) and centred on 400, 
420, 440 and 460 nm. 
The camera and filters are controlled by a local 
computer\footnote{http://beagleboard.org/}
and are housed in an aluminum box
mounted close to the centre of the VERITAS reflector 
in place of one of the mirror facets, as shown in Figure~\ref{box-mount}. 
Use of the standard facet-mounting hardware permits easy adjustment of the 
camera's pointing direction.

\begin{figure}[h]
  \centering
  \includegraphics[width=0.45\textwidth]{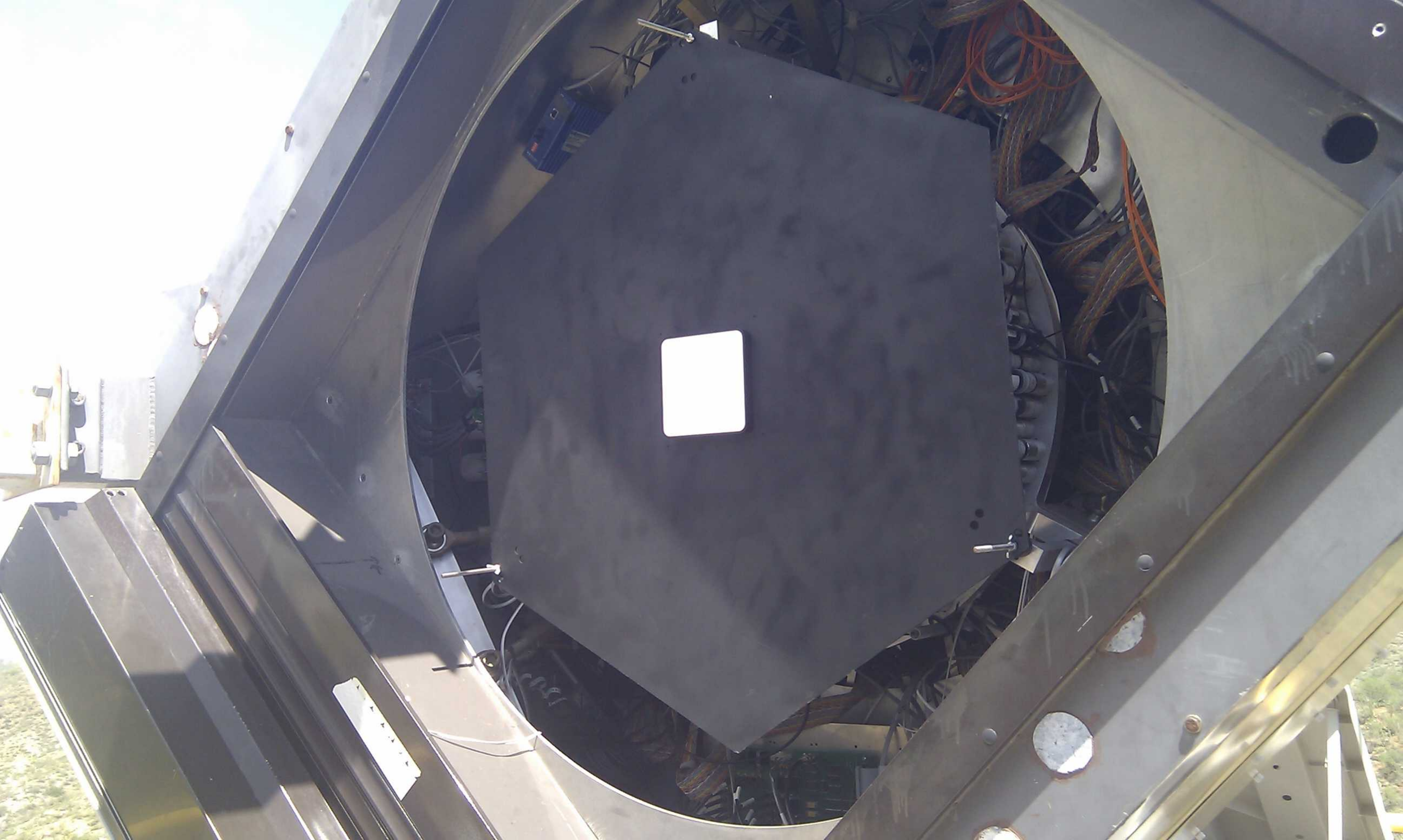}
  \vskip 0cm
  \caption{The Spectralon  target (white square)  
onto which the starlight is focussed is 
temporarily mounted at the focus of the telescope.}
  \label{target}
 \end{figure}

The secondary reflector onto which starlight is focussed is a square piece of
Spectralon \footnote{http://www.labsphere.com/}, 12.5 cm on a side.
It is a fluoropolymer with diffuse reflectance greater than 99\%
over the wavelength range of interest. 
The target is attached to an aluminum plate that can be temporarily mounted
at the focal point of the telescope, as shown in Figure~\ref{target}.

\section{Data Acquisition and Analysis}

Data for reflectivity studies are typically acquired over the course of several
nights, one night per telescope, at intervals of several months. 
This schedule is sometimes modified by weather and observing priorities.
For a given telescope, a star is tracked and images are recorded, first with
no filters in place and then with the filters, sequentially.
Exposure times are adjusted to avoid saturation, and multiple images are 
acquired to improve statistics. 
This sequence is repeated on a series of bright target stars at high elevation.

\begin{figure}[h]
  \centering
  \includegraphics[width=0.45\textwidth]{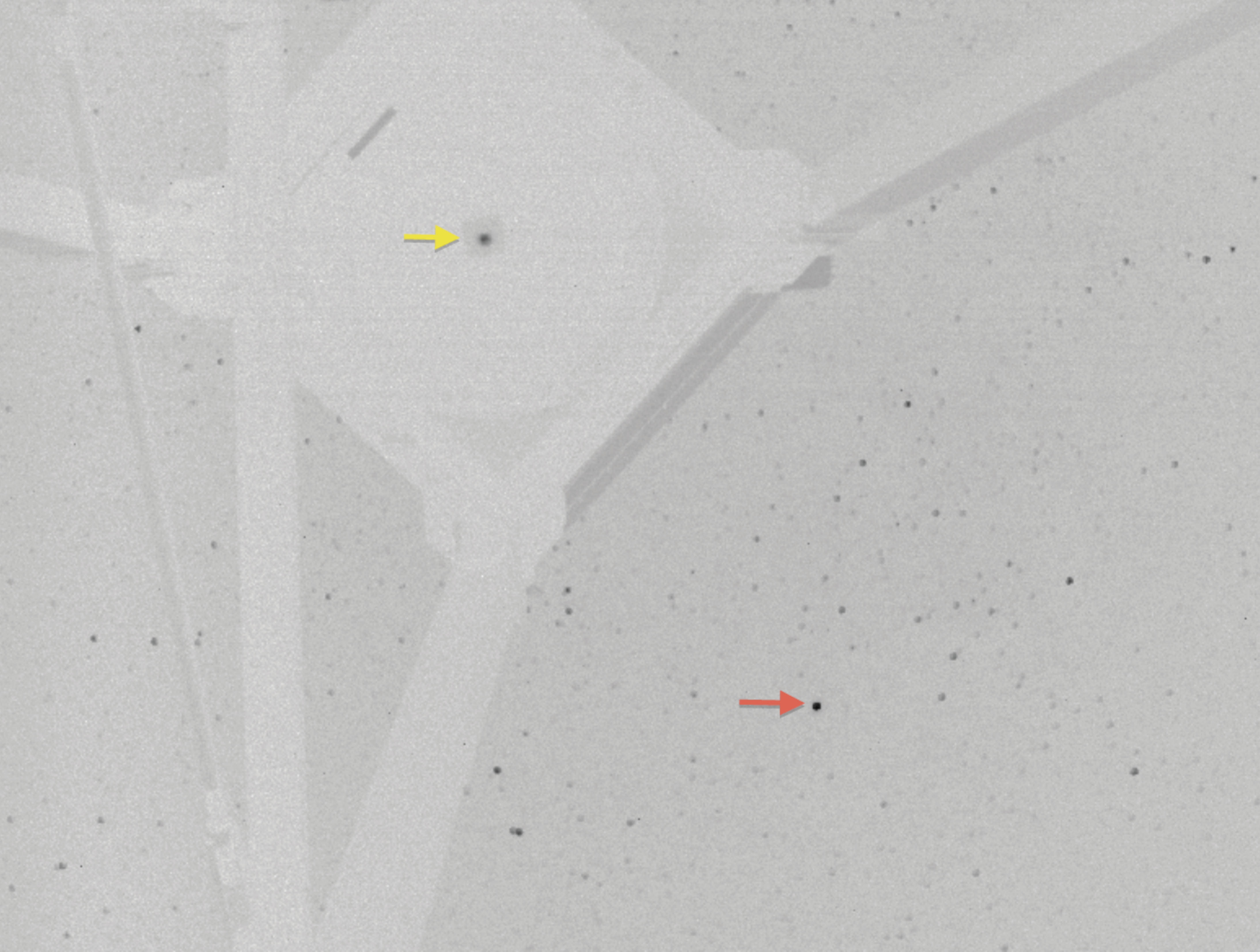}
  \vskip 0cm
  \caption{A sample image, typical of the many that are used in this study.
The telescope structure is clearly visible, illuminated by low-level ambient
light, and the Spectralon  target can be seen (yellow arrow), along with the glow from the
reflected star image. 
The direct light from the star forms the bright spot in the lower-right 
quadrant of this image (red arrow).
The intensity scale is logarithmic. 
}
  \label{sample_pic}
 \end{figure}

A typical image is shown in Figure~\ref{sample_pic} with a logarithmic 
intensity scale.
One can see the Spectralon  target and the broad glow of the reflected star
image in the upper left quadrant.
The bright spot due to the 
direct light from the target star is visible in the lower right quadrant.

The analysis proceeds by identifying the approximate locations of the 
star and the reflection in the image. 
The pixel values from a 100-by-100 square centred on each of these locations
are saved in files for further processing.
Files from multiple exposures are combined.

\begin{figure}[h]
  \centering
  \includegraphics[width=0.45\textwidth]{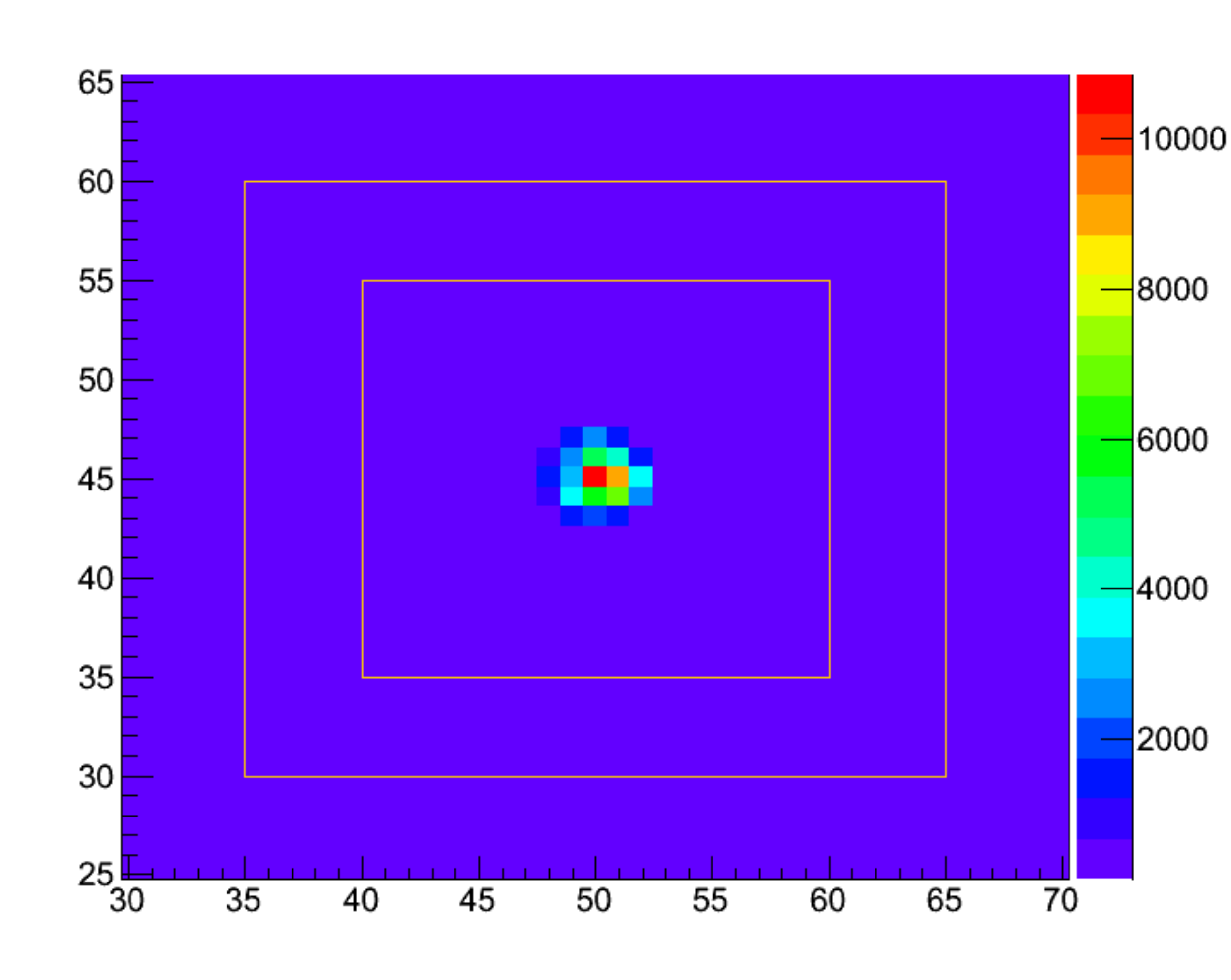}
  \vskip 0cm
  \caption{Two-dimensional histogram of pixel values from a 40-by-40 array
approximately centred on the position of the target star in the CCD image.
Background is estimated by averaging the values of pixels in the region between
the squares and is subtracted from the region within the inner square.
The sum over background-subtracted pixels from the inner square is used as the
signal from the target star.}
  \label{star}
 \end{figure}

A two-dimensional histogram of a typical star file is shown in 
Figure~\ref{star}, highlighting the good signal-to-noise ratio. 
The net signal is determined by summing the pixel values from the region
between the concentric squares and subtracting it, 
appropriately scaled, from the sum of the pixels from the inner square.
(The squares are overly large for this example but were defined such that 
they could be used for images where the star was defocussed to mitigate
saturation effects. 
We maintain the same limits for purposes of minimizing systematic errors.)
For the star image it is mainly the average pedestal that is being subtracted
since there is very little background light in that part of the image.

\begin{figure}[h]
  \centering
  \includegraphics[width=0.45\textwidth]{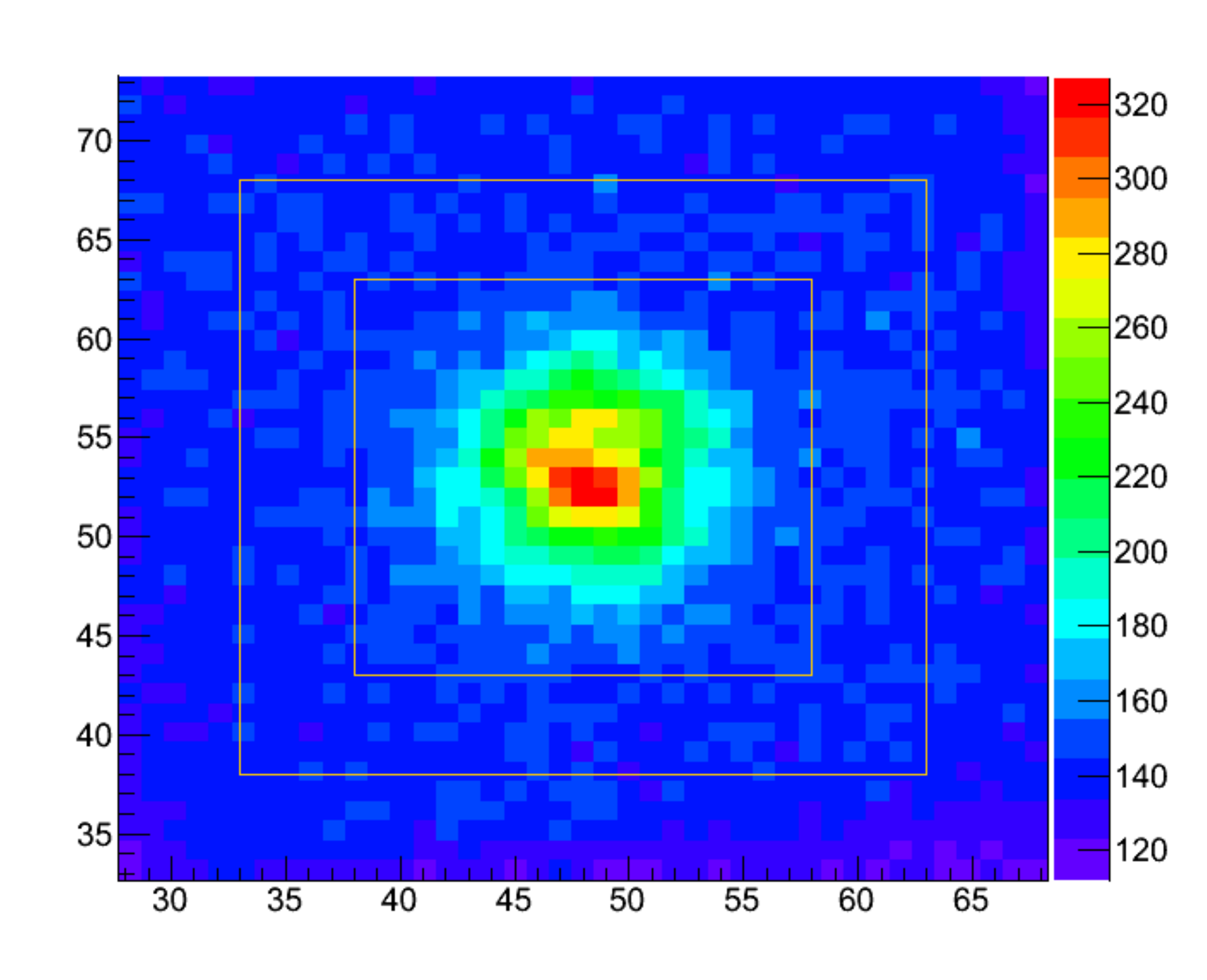}
  \vskip 0cm
  \caption{As in Figure~\ref{star} but for the reflection image.}
  \label{reflection}
 \end{figure}

\begin{figure}[h]
  \centering
  \includegraphics[width=0.45\textwidth]{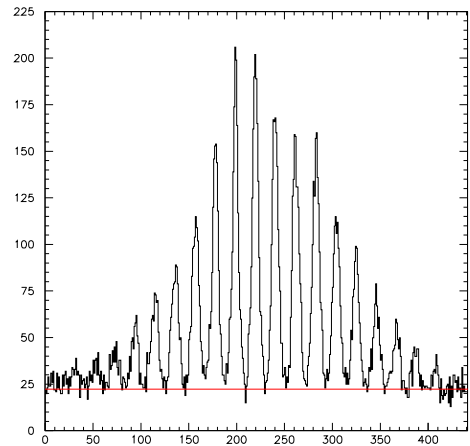}
  \vskip -2cm
  \caption{Pixel values from the inner square in Figure~\ref{reflection} 
plotted row by row together with the background estimate (red line). The
fact that the ``valleys'' reach, but do not go below, 
the background level and the tails at the left and right are approximately at
this level shows that background subtraction is done correctly.}
  \label{projection}
 \end{figure}

The corresponding plot for the reflection image 
is shown in Figure~\ref{reflection}
where it is clear that background is an important consideration.
Ambient light tends to brighten the Spectralon .
To check that background subtraction is done effectively and correctly we plot,
in Figure~\ref{projection}, the pixel values, row by row in sequence, 
from the inner square, together with a line indicating the computed 
background level.
The squares used here are based on 5- and 7-sigma limits determined by fitting 
a Gaussian distribution to the reflection distribution. 

The whole-dish reflectivity $R$ is calculated from the direct and reflected signals, $S_d$ 
and $S_r$, respectively, using the 
formula $ R = (S_r / S_d) \pi d^2 / A_M $ where 
$A_M$ is the area of the dish and $d$ is the distance from 
the Spectralon to the CCD camera.

\section{Corrections and Uncertainties}

\begin{figure}[h]
  \centering
  \includegraphics[width=0.45\textwidth]{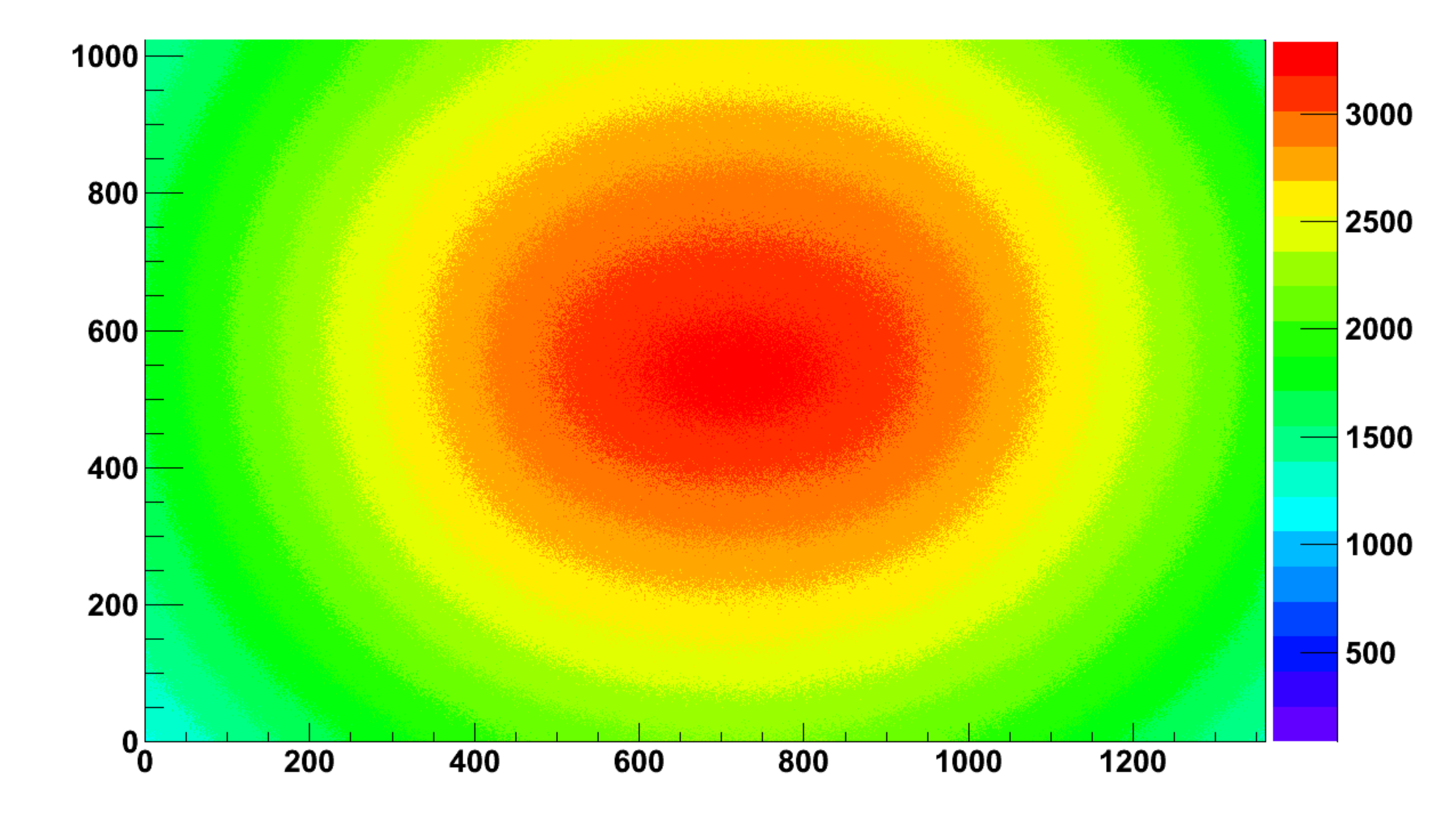}
  \vskip 0cm
  \caption{A flat-field image taken by pointing the camera to zenith on a 
clear night at twilight. The colour code shows the effects of vignetting.
Corrections on the order of 10\% are calculated from this image and applied 
to reflection values computed from images like that in Figure~\ref{sample_pic}.} 
  \label{flat}
 \end{figure}

The optics of our camera are such that vignetting is significant and must be 
taken into account when computing the reflectivity. 
We can estimate the size of the effect using images taken for flatfielding 
purposes.
These were acquired by pointing the camera to zenith at twilight and a sample 
is shown in Figure~\ref{flat}. 
The corrections can be determined by normalizing the pixel values to the 
maximum value and using the inverse of these numbers as pixel-by-pixel 
correction factors. 
In this work we use a single value for the star and a single value for the 
reflection since the variation is quite small over the regions of interest. 
The correction to the ratio of the reflection signal to the star signal is 
of order 10\%.

\begin{figure}[h]
  \centering
  \includegraphics[width=0.45\textwidth]{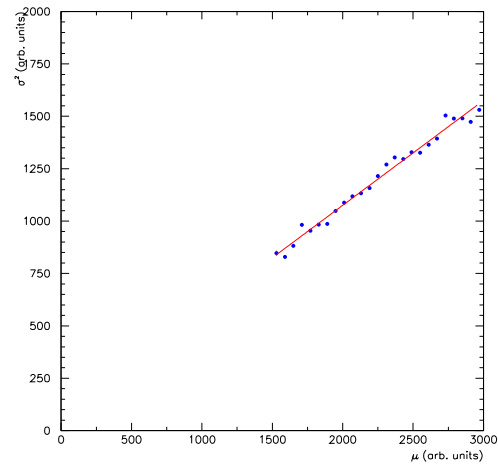}
  \vskip -2cm
  \caption{Pixel variance vs pixel value for distributions made with data from 
a flat-field image. The behaviour, $\sigma^2 = 60 + 0.51\mu$, 
where $\mu$ is the 
average value of neighbouring pixels, is used in determining statistical 
uncertainties in the computed reflectivities.}

  \label{error}
 \end{figure}

The data from Figure~\ref{flat} are useful for determining statistical 
uncertainties.
We assume that adjacent pixels should report approximately 
the same number in a flat-field
image since the twilight sky is uniform and vignetting effects vary slowly.
The largest contribution to any difference between adjacent values should be due
to statistical fluctuations. 
Thus we can histogram the difference between pixel $n$ and pixel $n+1$ for all 
values of $n$ and use the width of the resulting distribution as an estimator
of the statistical error.
This is expected to vary with the magnitudes of the pixel values so we make 
separate histograms for diffent ranges of pixel values. 
The results are shown in Figure~\ref{error} where the variances are 
plotted as a function of pixel value.
This dependence is used in assigning uncertainties to pixel values in the 
analysis.

\section{Some Results}

\begin{figure}[h]
  \centering
  \includegraphics[width=0.55\textwidth]{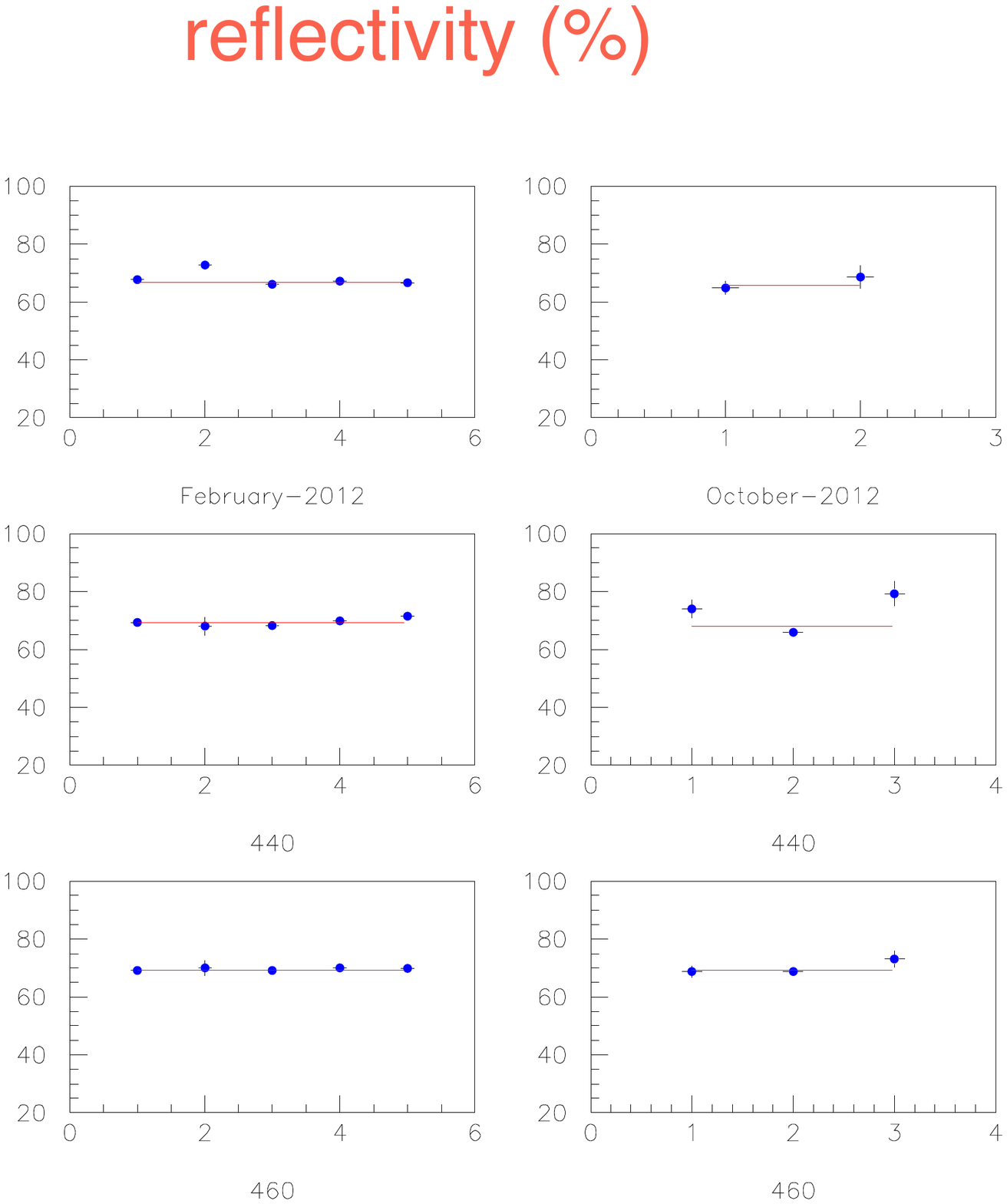}
  \vskip -2cm
  \caption{Reflectivity 
results, in per cent, for one of the VERITAS telescopes. 
The left-hand panels show data from February, 2012 and the right-hand panels 
from the following October. Data in the top panels were acquired without 
filters while those in the middle and bottom panels were acquired with 
band-pass filters at 440 nm and 460 nm, respectively. Each point within a 
panel was made using a different star. The flat red lines display the weighted
mean of the points.}
 \label{summary}
 \end{figure}

Some typical results, from one of the telescopes,
are shown in Figure~\ref{summary} where reflectivities in percent
are plotted.
Each point was calculated with images from a different star and the flat line
is an average value, drawn to show that results are consistent within errors. 
There are two columns, one for data acquired in February, 2012 and the other
for data acquired the following October. 
In each column the top panel shows reflectivities measured without a 
band-pass filter while the middle and bottom panels show measurements made with
440 nm and 460 nm filters, respectively. 
Images made with 400 nm and 420 nm filters had very poor signal to noise and 
are not shown. 
Measurements at these wavelengths will need longer integration times under 
darker skies.

Although the results shown are preliminary and we do not present precise 
numbers here, we note some interesting qualitative features.
First, the technique produces numbers that are reproducible, independent of
the star used, and stable over time.
Second, the reflectivities measured without a filter are slightly lower than
those measured with filters. 
This is presumably due to the fall-off with wavelength of reflectivity.
By restricting observations to wavelengths near the maximum of reflectivity 
curve one obtains larger values.
 
The whole-dish reflectivity numbers are in the neighbourhood of 70\%.
Laboratory measurements on individual facets before installation were reported
previously~\cite{emmet} and were in the range 90-95\% for wavelengths between
250 and 400 nm, falling to 75-80\% at wavelengths greater than 600 nm. 
These numbers are remeasured, and are found to be somewhat lower, when facets
are periodically removed for resurfacing, after which they return to their 
original values. 
The lower values for in situ measurements can be explained by features
such as shadowing of the mirror due to the PMT camera and support structure,
tails in the point-spread function due to residual misalignment of some of 
the mirror facets, and various degrees of surface aging - on any dish a certain
fraction of the facets will be due for resurfacing.
Indeed, the differences between laboratory-facet and whole-dish reflectivities 
are expected and are hard to calculate precisely so it is good to have an 
end-to-end system like the one described here.

\section{Conclusions}

The whole-dish reflectivity procedure suggested by MAGIC has been 
implemented in the VERITAS array and initial results indicate that it is 
relatively simple to use and produces reliable results.

As we progress towards a solid understanding of our measurements we intend to 
build a separate system for each telescope and take data in parallel. 
Similarly, with a better understanding of statistical and systematic 
uncertainties, we expect to reduce acquistion times to the point where they do 
not interfere significantly with astronomical observations.
It is important to have measurements on a monthly basis to be able to 
effectively track long-term changes. 
Measurements made immediately before and after mirror washings and facet 
replacements would also be of great interest.

\vspace*{0.5cm}
\footnotesize{{\bf Acknowledgments:}{
We warmly thank our colleagues in the VERITAS collaboration for their support
of this work and for assistance with data acquisition.
VERITAS research is supported by grants from the U.S. Department of Energy Office of Science, the U.S. National Science Foundation and the Smithsonian Institution, by NSERC in Canada, by Science Foundation Ireland (SFI 10/RFP/AST2748) and by STFC in the U.K. We acknowledge the excellent work of the technical support staff at the Fred Lawrence Whipple Observatory and at the collaborating institutions in the construction and operation of the instrument. }}

\end{document}